\documentclass[iop]{emulateapj}

\usepackage{lineno,hyperref,natbib}
\modulolinenumbers[5]

\def\aap{A\&A}
\def\mnras{MNRAS}  
\def\apjl{ApJ} 
\def\apj{ApJ} 
\def\aj{AJ} 
\def\aapr{A\&A~Rev.}
\def\memsai{Mem.~Soc.~Astron.~Italiana}

\begin{document}


\title{Dust in the torus of the AGN unified model}

\author{Rachel E. Mason}
\address{Gemini Observatory, 670 N. A'ohoku Place, Hilo, HI 96720, USA}


\begin{abstract}

These proceedings are based on an invited review talk at the 7th Meeting on Cosmic Dust. The scope of the meeting was broad, covering dust-related topics in areas from comets to debris disks and high-redshift galaxies. This is therefore intended to be an accessible, introductory overview of the dusty torus of the AGN unified model, aimed at summarizing our current understanding of the torus and with some emphasis on the solid-state spectral features observed.

\end{abstract}



\section{Introduction}

Active galaxies come in many varieties - by some estimates, as many as 60. Fortunately, the unified model of active galactic nuclei \citep[AGN; e.g.][]{Antonucci93} exists to make sense of the AGN ``zoo''. This model proposes that all AGN are composed of the same basic ingredients: a supermassive black hole and accretion disk, surrounded by a toroidal mass of dust and gas that obscures the nucleus from some lines of sight while collimating the radiation that escapes. ``Type 1'' objects (those with very broad emission lines in their optical spectra) are those in which we have a direct view of the hot, fast-moving material close to the accretion disk. In ``type 2'' AGN the torus obscures this material, and we see only narrow emission lines in the optical spectrum.

The validity of the basic unified model in at least some AGN was elegantly demonstrated by the classic observations of \citet{Antonucci85}. Their detection of polarized broad lines in the type 2 Seyfert galaxy, NGC~1068, is exactly what would be expected if that galaxy hosts a type 1 nucleus that is hidden from direct view, and visible only through photons that are scattered into our line of sight (becoming polarized in the process). However, the orientation of an otherwise identical central engine is almost certainly only one of several factors that govern the observed differences between AGN. AGN probably evolve over time, as a deeply-buried active nucleus eventually consumes or expels the material that surrounds it \citep[e.g.][]{Sanders88}. The properties of the torus may \citep{Lawrence82} or may not \citep{Lawrence10} depend on AGN luminosity, and some obscuration arises not in the torus but on large scales in the host galaxy \citep{Goulding12}. It has also been proposed both that the torus is part of an inflow that governs accretion onto the black hole, and that it is part of an outflowing wind where new dust is formed (\S\ref{flow}). The donut-like torus of the simplest version of the unified model captures only a fraction of the complexity of the dusty material in active nuclei.

Understanding the dusty structures in AGN is important for several reasons. To give just one example, a central question in astrophysics is how star formation and black hole growth have shaped the galaxies we see today. Both of these phenomena often take place in dust-enshrouded regions, so to track star formation and black hole activity over cosmic time, we need to understand the emission of dust heated by star formation and by an AGN. Several groups have been performing increasingly detailed radiative transfer modeling of local AGN \citep[e.g.][]{Pier92,Granato94,Nenkova08b,RamosAlmeida09,Honig10b,Honig10a,Alonso-Herrero11,Stalevski12}, and those models are now being applied to AGN at high redshift \citep{Deo11,Leipski14}, for which the observational data are usually much more limited\footnotemark.

\footnotetext{For a review of torus models, see \citet{Hoenig13}.}

In these proceedings I first summarize some of the fundamental characteristics of the torus, as currently understood. I then give a brief overview of theoretical work aimed at understanding why the torus exists, and how it might interact with its surroundings and change over time. Finally, I review the solid-state spectral features observed in the torus, point out what can be learned from them, and identify some deficits in our current understanding. 

\section{Essential facts about the torus}
\label{facts}

First, the torus is {\bf small}. It has never been resolved at the $\sim 0^{\prime\prime}.3$ angular resolution available from large, ground-based telescopes in the mid-IR, even in the nearest AGN. For example, \citet{Packham05} observed the Circinus galaxy, a type 2 Seyfert at a distance of only 4 Mpc, at 8.7 $\mu$m with $0^{\prime\prime}.33$ resolution. These observations set an upper limit of r$<$2 pc for the torus in that object. The exquisite resolution enabled by near- and mid-infrared interferometry is necessary to actually resolve the emission from the torus, and has revealed radii of $<$1 -- a few pc \citep{Jaffe04,Burtscher13}. Near-IR interferometry and reverberation mapping shows that the inner radius of the torus scales approximately as $L^{0.5}$ \citep{Suganuma06,Kishimoto11}, as expected if the inner boundary of the torus lies at the dust sublimation radius \citep{Barvainis87}. 

 Second, the torus is {\bf clumpy}. This is implied by several lines of evidence, perhaps most intuitively by the X-ray eclipse events observed in several AGN. \citet{Markowitz14} searched the Rossi X-Ray Timing Explorer archive for AGN observed at multiple epochs and showing changes in luminosity and column density that would indicate the passage of a discrete cloud across the line of sight. They found 12 such events in 8 objects, and were able to estimate the distances of the clouds from the black hole. In most cases the eclipsing clouds appeared to lie at the dust sublimation radius (i.e., the inner edge of the torus), and at similar distances to those inferred for dusty clouds through IR interferometry and reverberation mapping. This strongly suggests that the torus is composed of discrete, dusty clouds. Mid-IR interferometry also suggests that the torus is clumpy \citep{Tristram07}.

Clumpiness in the torus has some interesting implications \citep[for a review, see][]{Elitzur07}. Clumpiness breaks the strict relationship between orientation and observed AGN properties expected in the simplest version of the unified model. For example, even if an AGN is viewed from a ``face-on'' (polar) direction, if a single cloud happens to block the broad line region (BLR) from view, the object will be optically classified as a type 2. Clumpiness also means that dust temperature is not a simple function of distance from the AGN; individual clouds have hot, directly illuminated faces, and cool, obscured sides facing away from the accretion disk\footnotemark. The observed IR spectrum is the combination of the emission and absorption of all the clumps in the torus, and it has a weaker dependence on viewing angle than expected from models in which the torus is a uniform, homogeneous medium \citep{Buchanan06,Levenson09}.

\footnotetext{Measurements of the size of the torus over a range of wavelengths and extending into the sub millimeter could therefore provide an interesting probe of the detailed structure of the torus. Early ALMA observations of NGC~1068 at $\sim 0^{\prime\prime}.3 - 0^{\prime\prime}.5$ resolution have not resolved the torus \citep{Garcia-Burillo14}, however, so its size at submillimeter wavelengths is not well constrained at the time of writing. }

Besides small-scale clumpiness, the torus appears to be {\bf structured} on larger scales as well.  \citet{Tristram14} performed extensive mid-IR interferometric observations of the Circinus galaxy, providing unusually good sampling of the {\it uv}  plane. The data suggest that the IR-emitting material is distributed in the form of a disk orientated perpendicular to the ionization cone, and a more extended component elongated along the polar direction. The polar, extended component is interpreted as the inner funnel of a more extended structure that is responsible for most of the obscuration and collimation of the nuclear radiation: the torus. The disk is presumably related to the warped maser disk observed in this galaxy \citep{Greenhill03}, while a dense disk and outer, diffuse structure have been seen in some simulations of the torus (\S\ref{flow}).
 
Finally, the torus is a {\bf dynamic, changing} entity. As we will see in \S\ref{flow}, it may be composed of material that is flowing towards and/or away from the black hole. There is also evidence that the torus is a site of dust destruction and re-formation. \citet{Kishimoto13} present estimates of the inner radius of the torus in NGC~4151, derived from reverberation mapping and interferometry and dating back to the 1970s. They compare this with historical measurements of the AGN luminosity, and find an increase in torus radius following a flare from the AGN. The radius has remained large as the AGN has returned to its low state. Kishimoto et al. interpret this as evidence of dust being destroyed by the increased emission from the central engine, and predict a reformation timescale of several years.

\section{Inflow, outflow, and the origin of the torus}
\label{flow}

What mechanism(s) create the torus, and how might it evolve over time? Several recent theoretical studies have addressed these issues. In a series of ``nested'' simulations, \citet{Hopkins12} follow gas inflow from scales of $\sim$100 kpc to $<$1 pc from the nucleus. The inflow from Galactic scales creates eccentric disks of gas and stars with radii of $\sim$0.1 -- 10 pc. These disks contain clumps, warps and twists, and exert torques on the gas that drive further inflow. Their sizes, clumpy nature, gas masses, column density distributions,  etc., are comparable to expectations for the torus, leading Hopkins et al. to identify them with the torus itself. In this scenario, therefore, the torus strongly influences accretion of matter by the black hole. 

In the \citet{Hopkins12} simulations, nuclear star formation is a consequence of the gas inflow that forms the torus. In the simulations of \citet{Schartmann09}, the torus is a consequence of nuclear star formation. Stellar mass loss leads to filamentary inflows, creating a diffuse structure on scales of 10s of pc. On smaller scales, $\sim$1 pc, a dense, turbulent disk is formed. In the specific case of NGC~1068, \citet{Schartmann10} find that the predicted gas masses, column densities, and disk sizes agree quite well with the observations.

Alternatively, the torus may be composed of material flowing {\em outwards} from the nucleus. \citet{Elitzur06} examined the possibility that the torus is a region in a hydromagnetic wind flowing off the accretion disk \citep[e.g.][]{Emmering92} in which conditions are suitable for dust to form. If the dust becomes optically thick, it will obscure the nucleus in edge-on views while leaving it visible from the polar direction. One corollary of this model is that, during periods when little material is flowing into the nucleus, a long-term outflow cannot be sustained. In low-accretion-rate objects, then, the torus should cease to exist. Whether this actually happens is not yet clear. On the one hand, \citet{Elitzur09} show that the BLR is absent in AGN with luminosities $< 5 \times 10^{39} (\rm M_{BH}/10^7 M_\odot)^{2/3} \rm \; erg \; s^{-1}$. In the disk wind model the BLR is simply the region of the wind inside the dust sublimation radius, so if the BLR is absent then it follows that the torus will be as well. However, the near- to mid-IR SEDs of low-luminosity AGN do not differ clearly from those of higher-luminosity objects, suggesting that the torus persists at least in the objects studied so far \citep{Mason12,Mason13}.

Could AGN outflows be an important source of dust production at high redshift? Already, by z$\sim$6, quasars contain as much as 10$^8 \rm M_\odot$ of dust \citep{Bertoldi03}. The presence of this much dust at such early times implies a very rapid formation mechanism. Considerable attention has been given to dust production by short-lived, massive stars and their supernovae \citep[e.g.][]{Gall11}, while AGN outflows have been less well studied. \citet{Elvis02} modeled the production of dust in expanding BLR clouds, noting that the conditions are expected to be similar to those encountered in dust-forming stellar winds. They estimated that 0.01M$_\odot$/yr of dust can be produced in quasar winds, and argued that some of this will be ejected into the intergalactic medium. \citet{Maiolino06} suggest that this source of dust could be responsible for all of the dust observed in SDSSJ1148+52, a quasar at z=6.4, while \citet{Pipino11} calculate that AGN winds will be important only at early times and in the central regions of galaxies. All of these authors acknowledge that their results are sensitive to the model assumptions, which are in many cases not well constrained. Direct observational evidence of dust being produced in AGN winds would therefore be very interesting.

In practice, inflow, outflow and star formation probably all play complex and interconnected roles in shaping the nuclear obscuration in AGN.  Further simulations \citep[e.g.][]{Wada12} and observations aimed at elucidating these roles will be important in shaping our understanding of the nature of the torus and its effect on accretion and the host galaxy.

\section{Solid-state spectral features in the torus}
\label{dust}

A number of solid-state spectral features are observed in galaxy nuclei. They include:

\begin{itemize}
\item Silicate features, seen in emission or absorption around 10 and 20 $\mu$m;
\item The 3.4 $\mu$m C-H stretch, produced by aliphatic hydrocarbons in dust grains and observed in absorption;
\item Polycyclic aromatic hydrocarbon (PAH) emission features, such as those at 3.3, 6.2, 7.7, 8.6, 11.3, and 12.7 $\mu$m;
\item Ice absorption bands, including the 3.0 and 6.0 $\mu$m H$_2$O features among others.
\end{itemize}

The ice features are observed in objects like UltraLuminous Infrared Galaxies (ULIRGs), whose active nuclei are usually deeply buried in dust \citep{Spoon02}. These objects may represent an early stage in the evolution of an active galaxy, after the onset of star formation and AGN activity, but before the nucleus has escaped from its cocoon of gas and dust. In fact, \citet{Spoon02} speculate that the relative strengths of ice absorption and PAH emission in their ULIRG sample may reflect a sequence from strongly obscured nascent star formation/AGN activity to a less obscured, later phase.

PAH emission features are commonly observed in the central few hundred parsecs of active galaxies \citep[e.g.][]{Imanishi04b,Smith07}. However, at smaller distances from the central engine they tend to be weak or undetectable \citep{Honig10a}. This could be because of destruction, dilution, or suppression of star formation. Based on laboratory experiments, \citet{Voit92} predict that the fragile PAH molecules/grains will be destroyed by X-rays from the central engine of AGN, unless they exist in highly shielded regions (e.g. within the torus). Even if PAH molecules are abundant, it may be difficult to detect their spectral features in the centers of galaxies where the AGN continuum emission can be strong. However, \citet{Esquej14} find no relationship between PAH equivalent widths and AGN luminosity in the nuclear regions of their sample of 29 Seyfert galaxies. They point out that although the PAH fluxes imply a lower absolute star formation rate (SFR) in the nucleus than on larger scales, the SFR density in the nucleus is actually much higher.

The 3.4 $\mu$m absorption results from stretching of C--H bonds in aliphatic hydrocarbon dust grains. It is ubiquitous in the diffuse interstellar medium of our galaxy, but has not been detected in dense molecular clouds \citep[e.g.][]{Pendleton94,Chiar96}. The feature is weak compared to the 10 $\mu$m silicate feature, but it has been detected in several Seyfert 2 galaxies and ULIRGs with large obscuring columns. In particular, it is detected in NGC~1068 \citep{Imanishi97,Mason04}. This is a nearly face-on spiral galaxy, which implies that the 3.4 $\mu$m feature comes from the nucleus and not the large-scale ISM of the host galaxy. The profile of the 3.4 $\mu$m feature in other galaxies matches very closely that observed in the Galactic diffuse ISM, suggesting similar formation and processing pathways \citep{Mason04,Dartois04}. This, along with the non-detection of CO absorption bands in NGC~1068 \citep{Lutz04,Mason06,Geballe09}, has led to the suggestion that the dust in the torus may more closely resemble the Galactic diffuse ISM than typical molecular clouds \citep{Geballe09}.

The silicate features have been very important to our understanding of dust in AGN. Na\"{i}vely, one would expect to observe silicate emission in type 1 AGN (in which we should have a direct view of hot dust in the inner ``funnel'' of the torus), and absorption in type 2s (where the hot dust is obscured by a large column of cool dust; see the cartoon in \citet{Kohler10}). Silicate {\em absorption} has been known for decades \citep[e.g.][]{Lebofsky79}, and the survey of \citet{Roche91} showed that it is widespread. {\em Emission} features, on the other hand, were elusive. The failure to detect them led to suggestions that silicates must be depleted in the inner torus, or that the grain size distribution in the inner torus is biased towards large grains. With the advent of the Spitzer Space Telescope, however, strong silicate emission features were discovered in quasars and Low Ionisation Nuclear Emission Region galaxies \citep[LINERs; ][]{Hao05,Siebenmorgen05,Sturm05}. While this solved the problem of the ``missing" silicate emission, the observations began to raise some new questions. First, the silicate emission features have now been detected not only in type 1 AGN, but also in type 2 objects \citep{Hao07}. Second, even the original discovery papers noted that the emission feature profiles are broadened and shifted to longer wavelengths than the features observed in the Galactic ISM.

The detection of silicate emission in type 2 AGN may be related to the clumpiness of the torus which, as noted in \S\ref{facts}, breaks the strict relation between orientation and AGN properties. Using the clumpy radiative transfer models of \citet{Nenkova08b}, \citet{Nikutta09} found that the observed distribution of silicate feature strengths (in emission and absorption) among AGN types can be explained by clumpiness. In modeling the SED of a type 2 quasar with the 10 $\mu$m silicate feature in emission, they note that the appearance of the feature in emission simply requires a direct view of enough of the hot clump faces on the far side of the AGN.  \citet{Mason09} were also able to reproduce the observed spectrum of a type 2 Seyfert with silicate emission by invoking a ``slender'' clumpy torus.

Alternatively, it has been suggested that the silicate emission is produced by dust in the narrow-line region (NLR), rather than the torus \citep{Schweitzer08}.  At the time of writing, high-resolution, spatially-resolved mid-IR spectroscopy exists for only one object, NGC~1068 \citep{Mason06,Rhee06}, which is a type 2 AGN with the silicate feature in absorption, not emission. If future observations detect silicate emission in an extended NLR, this would have to be taken into account when deriving torus parameters from unresolved IR spectroscopy. 

Many explanations have been proposed to account for the unusual profiles of the silicate features in emission (and, occasionally, in absorption). \citet{Nikutta09}, for example, argue that radiative transfer effects are responsible for the broad, redshifted silicate emission in PG1211+143. In their model, the broadening of the feature is due to the peak of the emission being absorbed by an intervening cloud. The wavelength shift is caused by the underlying, rising continuum. The broad, redshifted feature in the LINER M81 has been successfully modeled using large ($\sim$5 $\mu$m), porous, amorphous olivine and carbon grains \citep{Smith10}. In NGC~1068, the absorption feature observed in interferometric data (i.e., on milli-arcsec or $\sim$parsec scales) is also anomalous, with absorption starting at longer wavelengths than observed in the Galactic ISM. Various chemical compositions have been suggested to account for the unusual profile: while \citet{Jaffe04} obtain a good fit using Ca$_2$Al$_2$SiO$_7$ (``gehlenite"), \citet{Kohler10} favor SiC. Interesting dust chemistry has also been implicated in the silicate emission features in PG2112+059, a broad absorption line (BAL) quasar. \citet{Markwick-Kemper07} fit the mid-IR spectrum of this object with a mixture of species: amorphous olivine; crystalline forsterite; corundum (Al$_2$O$_3$); periclase (MgO); and a minor contribution from PAHs. 

Clearly, there are many possible explanations for the anomalous silicate feature profiles in AGN. It is less clear what the implications of each might be. The Al$_2$O$_3$ used to model the features in PG2112+059 is a dust precursor molecule, so its possible presence in this BAL quasar could be relevant to the question of dust production in AGN winds in the early universe (\S\ref{flow}). If non-standard dust properties are necessary to explain the features, the appropriate kind of dust should in principle be incorporated into the models that predict the IR spectra of AGN. Whether this would significantly affect  the conclusions drawn from the models, though, has not yet been thoroughly investigated. In any event, as most detailed analysis of the features has so far concentrated on a handful of individual galaxies, the field would benefit from a systematic survey of the profiles of the silicate features in AGN.

\section{Conclusions}

As the cornerstone of the AGN unified model, the dusty torus has been the subject of much research in the last few decades. 
Aided by advances in observational and computational capabilities, we are now starting to appreciate its complexity. The community is beginning to consider the torus less as a well-defined, isolated structure and more in the context of its possible interactions with its surroundings. Many open questions remain, particularly concerning the origin of the torus and how it changes over the lifecycle of an AGN. A closer examination of the roles of inflows and outflows in shaping the torus could produce some valuable insights in these areas.

\vspace*{2mm}

Supported by the Gemini Observatory, which is operated by the Association of Universities 
for Research in Astronomy, Inc., on behalf of the international Gemini partnership of 
Argentina, Australia, Brazil, Canada, Chile, and the United States of America.


\end{document}